\documentstyle[twoside,fleqn,espcrc2,epsf]{article}
\newcommand{\lw}[1]{\smash{\lower2.ex\hbox{#1}}}
\def\simlt{\rlap{\lower 3.5 pt\hbox{$\mathchar \sim$}}\raise 1pt \hbox {$<$}}
\def\simgt{\rlap{\lower 3.5 pt\hbox{$\mathchar \sim$}}\raise 1pt \hbox {$>$}}

\title{Non-Perturbative Determination of $c_{\rm SW}$ in Three-flavor 
Dynamical QCD\thanks{Talk presented by S.~Aoki}
}

\author{CP-PACS and JLQCD Collaborations:
  S.~Aoki\rlap,\address{Institute of Physics,
    University of Tsukuba, Tsukuba, Ibaraki 305-8571, Japan}
  M.~Fukugita\rlap,\address{Institute for Cosmic Ray Research,
    University of Tokyo, Kashiwa, Chiba 277-8582, Japan}
  S.~Hashimoto\rlap,\address{High Energy Accelerator Research Organization
    (KEK), Tsukuba, Ibaraki 305-0801, Japan}
  K-I. Ishikawa\rlap,$^{\rm a,}$\address{Center for Computational Physics,
    University of Tsukuba, Tsukuba, Ibaraki 305-8577, Japan}
  N.~Ishizuka\rlap,$^{\rm a,d}$
  Y.~Iwasaki\rlap,$^{\rm a}$
  K.~Kanaya\rlap,$^{\rm a}$
  T.~Kaneko\rlap,$^{\rm c}$
  Y.~Kuramashi\rlap,$^{\rm c}$
  M.~Okawa\rlap,\address{Department of Physics, Hiroshima University, 
Higashi-Hiroshima, Hiroshima 739-8526, Japan}
  V.~Lesk\rlap,$^{\rm d}$
  Y.~Taniguchi\rlap,$^{\rm a}$
  N.~Tsutsui\rlap,$^{\rm c}$
  A.~Ukawa\rlap,$^{\rm a,d}$ 
  T.~Umeda\rlap,$^{\rm d}$
  N.~Yamada$^{\rm c}$ and
  T.~Yoshi\'e$^{\rm a,d}$
}
\begin{document}
\begin{abstract}
We present a fully non-perturbative determination of the $O(a)$ improvement
coefficient $c_{\rm SW}$ in three-flavor dynamical QCD for the
RG improved as well as the plaquette gauge actions, using
the Schr\"odinger functional scheme.
Results are compared with one-loop estimates at weak gauge coupling.
\end{abstract}

\maketitle

\section{Introduction}
\label{sec:intro}

Realistic simulation of QCD requires treating the light
up, down and strange quarks dynamically.  Incorporating a
degenerate pair of up and down quarks have become almost
standard by now, and a first attempt toward the continuum
extrapolation has shown that the deviation of the 
quenched hadron mass spectrum from experiment \cite{cppacsQ} 
is sizably reduced\cite{cppacsF}.
Adding a dynamical strange quark is the next step, which has become
possible with the recent algorithmic development for odd number of 
fermions\cite{jlqcdPHMC}.

The CP-PACS and JLQCD Collaborations have jointly started 
a 2+1 flavor dynamical QCD, employing the polynomial HMC
(PHMC) algorithm for strange quark and the HMC algorithm for
up and down quarks.  We choose the renormalization-group  (RG) improved
action for the gauge fields,  in order to avoid the lattice artifact
present for the plaquette action\cite{jlqcdPT}.
We wish to use a fully $O(a)$-improved action for quarks to control lattice
spacing errors.
Here we report on a non-perturbative determination of $c_{\rm sw}$ for
three-flavor QCD by the Schr\"odinger functional scheme both for the plaquette
and RG-improved gauge actions.

\section{Method and Simulations}
For the determination of $c_{\rm SW}$,
we basically follow the method of ref.\cite{JS},
except for the choice B for the boundary weight of the RG-improved gauge action
\cite{AFW}.
We refer to ref.~\cite{JS} and references therein for notations in this report.

\begin{figure}[bt]
%\vspace{-0.5cm}
\centerline{\epsfxsize=7.0cm \epsfbox{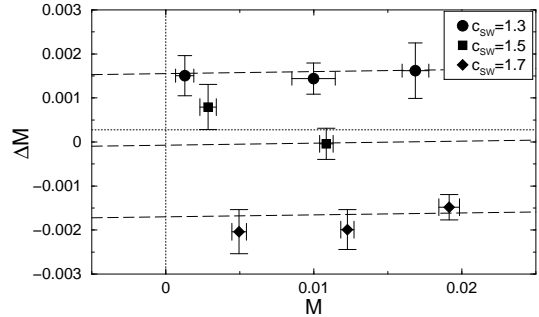}}
\vspace{-0.8cm}
\caption{$\Delta M $ as a function of $M$ for the RG action with $N_f=3$ 
at $\beta = 2.2$.}
\label{fig:dMM}
\vspace{-0.7cm}
\end{figure}
We mainly use an $8^3\times 16$ lattice in our determination of $c_{\rm SW}$
for the RG-improved as well as the plaquette action
with $N_f=3$ dynamical quarks at several values of $\beta$. 
Simulations with $N_f = 4,2,0$ are also made for comparison.

We  measure the modified PCAC quark masses, $M$ and $M^\prime$,
and their difference $ \Delta M = M - M^\prime $,
at several values of $c_{\rm SW}$ and $K$. 
We have taken these parameters to realize $M=0$ by an interpolation,
except at strong couplings for the case of $N_f=3$,
where an extrapolation to $M=0$ is necessary as shown
in Fig.~\ref{fig:dMM}.

From the linear fit of $\Delta M$ as a function of $M$ and $c_{\rm SW}$:
$
\delta M = a_0 + a_1 M + a_2 c_{\rm SW},
$
we obtain the $O(a)$ improvement coefficient
$
c_{\rm SW} = (\Delta M^{(0)} - a_0)/a_2
$,
where $\Delta M^{(0)} = 0.000277$, marked by the horizontal dotted line in 
Fig.~\ref{fig:dMM}, is the tree-level value of $\Delta M$ on
a $8^3\times 16$ lattice.
Note that the dependence of $\Delta M$ on $c_{\rm SW}$ becomes weaker
at stronger couplings, so that the determination of $c_{\rm SW}$ is 
more difficult, and hence the error is larger, at stronger couplings.

\section{Results}
\begin{figure}[bt]
\vspace{-0.5cm}
\centerline{\epsfxsize=6.5cm \epsfbox{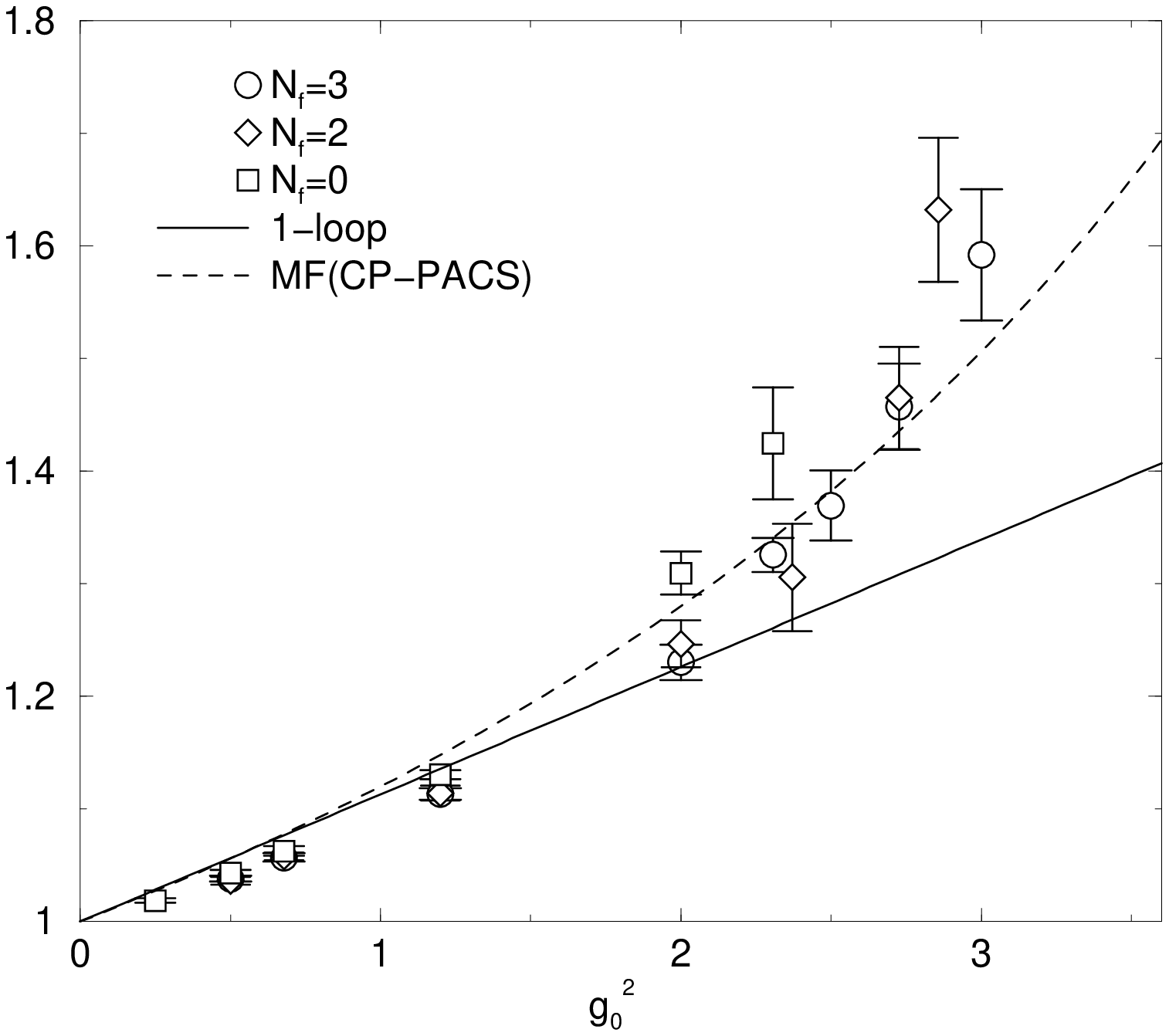}}
\centerline{\epsfxsize=6.5cm \epsfbox{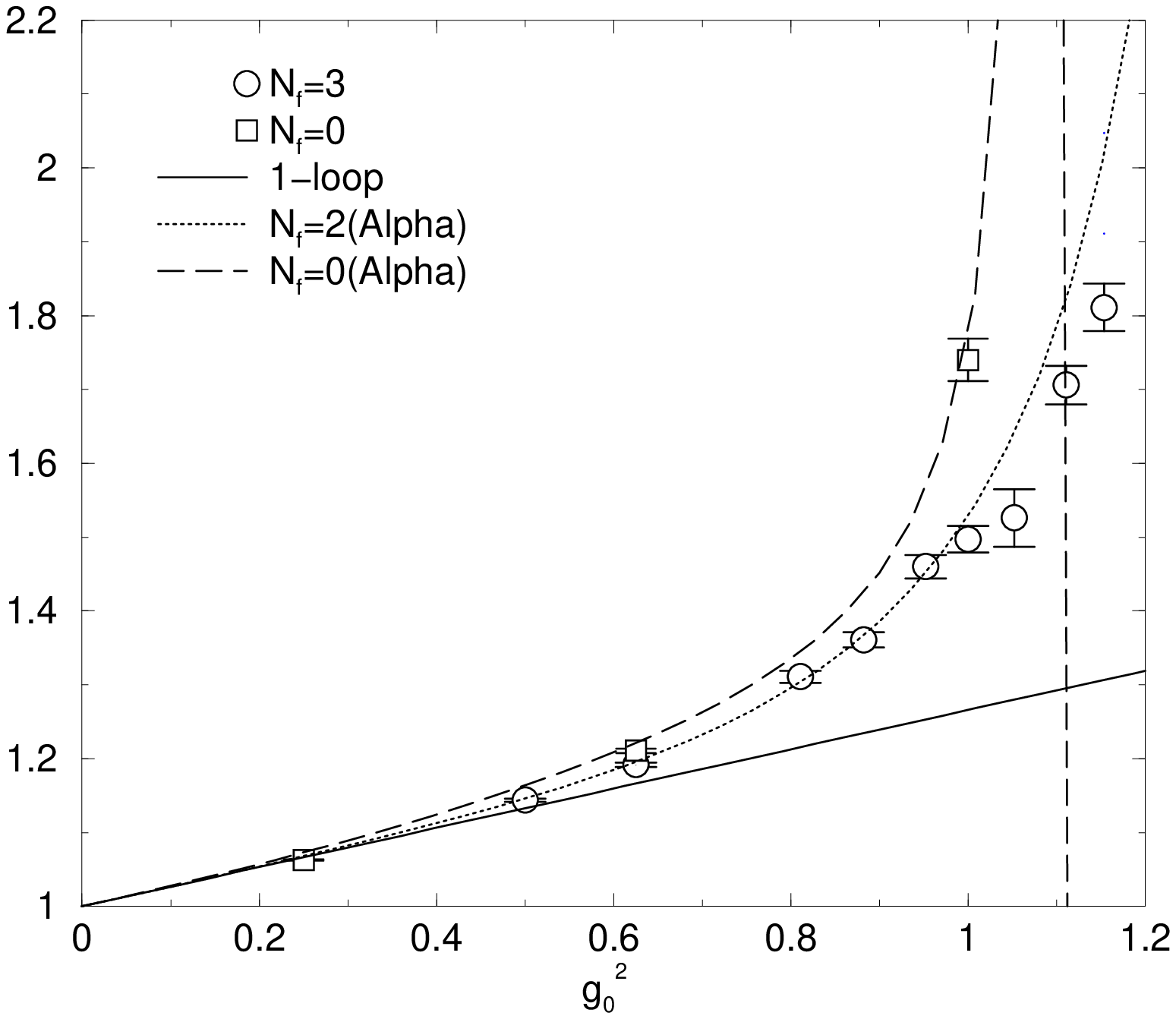}}
\vspace{-0.8cm}
\caption{$c_{\rm SW}$ as a function of $g_0^2$ for the RG action
(upper) and for the plaquette action(lower).}
\label{fig:cswRP}
\vspace{-0.7cm}
\end{figure}
In the upper plot of Fig.~\ref{fig:cswRP} 
we show the non-perturbative value of $c_{\rm SW}$
as a function of the bare gauge coupling $g_0^2$ 
for the RG-improved gauge action
with $N_f=3$(circles), 2(diamonds) and 0(squares),
together with the one-loop estimate(solid line) and the mean-field(MF)
estimate(dashed line) used in ref.~\cite{cppacsF}.
Similarly, results for the plaquette action with $N_f=3$(circles)
and $0$(squares) are given in the lower plot of Fig.~\ref{fig:cswRP},
together with the one-loop estimate(solid line) and
the non-perturbative values by the Alpha collaboration
with $N_f=2$(dotted lines)\cite{JS} and 0(long-dashed line)\cite{alphaCSW}.

\begin{figure}[bt]
\vspace{-0.5cm}
\centerline{\epsfxsize=6.5cm \epsfbox{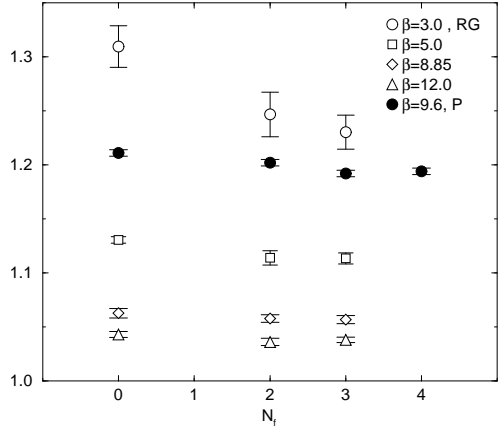}}
\vspace{-0.9cm}
\caption{$c_{\rm SW}$ as a function of $N_f$ for RG and  P(plaquette) 
actions.}
\label{fig:cswNF}
\vspace{-0.4cm}
\end{figure}

In both cases, the non-perturbative values of $c_{\rm SW}$ are almost
$N_f$ independent at weak coupling while they become larger for smaller $N_f$
at strong coupling. This tendency can be clearly seen in 
Fig.~\ref{fig:cswNF}, where
$c_{\rm SW}$ is plotted as a function of $N_f$ for the RG action(open symbols)
and the plaquette action(solid circles).

\section{Comparison with perturbative estimates}
\begin{figure}[tbh]
\vspace{-0.4cm}
\centerline{\epsfxsize=6.5cm \epsfbox{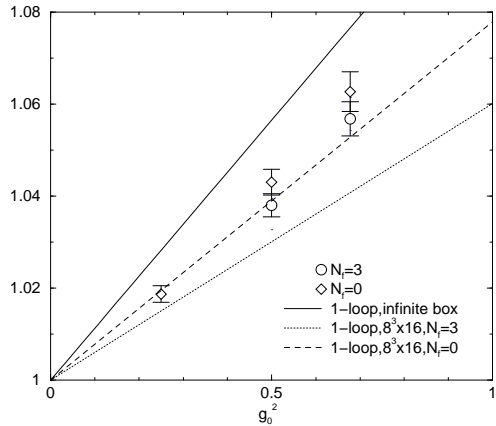}}
\vspace{-1.0cm}
\caption{$c_{\rm SW}$ for the RG action at weak coupling, together
with the one-loop estimate on the $8^3\times 16$ lattice
for $N_f=3$(dotted line) and $N_f=0$(dashed line).}
\label{fig:cswRG0}
\vspace{-0.8cm}
\end{figure}
At first sight, the non-perturbative $c_{\rm SW}$ seems to undershoot
the one-loop estimate at weak coupling for the RG action, while it
converges smoothly from above for the plaquette action. 
We have found that the discrepancy seen for the RG action is caused
by the one-loop  $O(a /L)$ error in $c_{\rm SW}$
\footnote{$O(a)$ errors of $c_{\rm SW}$ in general  cause $O(a^2)$ errors 
in on-shell quantities, which are irrelevant in the $O(a)$ improvement.},
which becomes leading 
after the $O(a/L)$ error at tree level is removed by requiring 
$\Delta M = \Delta M^0$.
In Fig.~\ref{fig:cswRG0}, the non-perturbative $c_{\rm SW}$ is 
compared with the one-loop estimate we have calculated 
on the same lattice size employed in the simulation, $8^3\times 16$. 
As seen from the figure the non-perturbative value agrees with the one-loop 
estimate much better on the $8^3\times 16$ lattice than in the infinite box.
Note that the $O(g_0^2 a/L)$ contribution to $c_{\rm SW}$ slightly depends 
on $N_f$ through the fermion tadpole in the presence of the background
gauge field of the Schr\"odinger functional scheme.
Such an $N_f$ dependence is indeed seen in the numerical data of
Fig.~\ref{fig:cswRG0}.

\begin{figure}[thb]
\vspace{-0.5cm}
\centerline{\epsfxsize=6.4cm \epsfbox{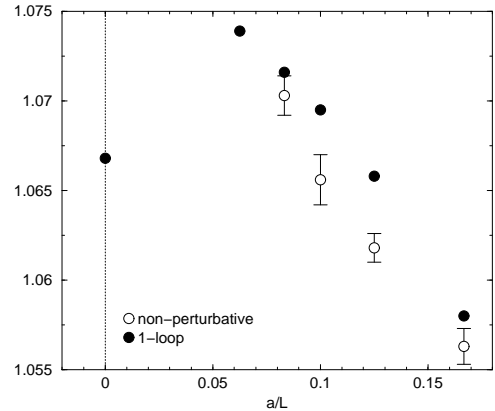}}
\vspace{-1.0cm}
\caption{Non-perturbative $c_{\rm SW}$(open circles)
and one-loop estimate(solid circles) 
as a function of $a/L$ at $\beta =24$ for the plaquette action with $N_f=0$.}
\label{fig:cswL}
\vspace{-0.8cm}
\end{figure}

For the plaquette action we also confirm a presence of the $O(g_0^2 a/L)$ 
correction, which is small on the $8^3\times 16$ lattice, 
as shown in Fig.~\ref{fig:cswL} where $c_{\rm SW}$ 
is plotted as a function of $a/L$ at $\beta =24$ with $N_f=0$. 
The one-loop estimates(solid circles) reproduce the non-monotonic
behaviour of non-perturbative values(open circles) well.

\section{Discussion}
We have determined the non-perturbative value of $c_{\rm SW}$ for the RG action
at several gauge couplings with $N_f=3,2,0$.
In order to obtain an interpolation formula of $c_{\rm SW}$ 
as a function of $g_0^2$, 
we have to eliminate large $O(g_0^2 a/L)$ corrections to $c_{\rm SW}$
present for the RG action. We are currently investigating this problem.

We are also measuring the hadron spectrum for the RG action
at $\beta \equiv 2$ with $N_f=3$ using the preliminary value
of $c_{\rm SW}$,
in order to determine the corresponding lattice spacing.

\vspace{2mm}
This work is supported in part by Grants-in-Aid of the Ministry of Education
(Nos.
11640294, % okawa
12304011, % iwasaki
12640253, % saoki
12740133, % ishizuka
13135204, % iwasaki
13640259, % ukawa
13640260, % kanaya
14046202  % saoki
14740173  % kaneko
).
N.Y. is supported by the JSPS Research Fellowship.


\begin{thebibliography}{99}

\bibitem{cppacsQ}CP-PACS Collaboration: S.~Aoki, {\it et al.},
Phy.~Rev.~Lett.~84 (2000) 238.

\bibitem{cppacsF}CP-PACS Collaboration: A.~Ali~Khan, {\it et al.},
Phy.~Rev.~Lett.~85 (2000) 4674

\bibitem{jlqcdPHMC}JLQCD Collaboration: S.~Aoki, {\it et al.},
 Phy.~Rev.~D65 (2002) 094507, and references therein.

\bibitem{jlqcdPT}JLQCD Collaboration: S.~Aoki, {\it et al.},
Nucl.~Phys.(Proc.Suppl.)~106 (2002) 263.

\bibitem{JS}K.~Jansen and R.~Sommer, Nucl.~Phys.~B530 (1998) 185.

\bibitem{AFW}S.~Aoki, R.~Frezzotti, P.~Weisz, Nucl.~Phys.~B540 (1998) 501.

\bibitem{alphaCSW}M.~L\"ushcer {\it et al.},
Nucl.~Phys.~B491 (1997) 323.

\end{thebibliography}
\end{document}